\documentclass[notitlepage,preprintnumbers,prd,longbibliography,showpacs,nofootinbib]{revtex4-1}
\usepackage{amsmath,amssymb}
\usepackage{graphics,color,subfigure}
\usepackage{epsfig}
\usepackage{dcolumn}% Align table columns on decimal point
\usepackage{bm}% bold math
\usepackage{longtable}

\usepackage[colorlinks=true,
            urlcolor=blue,
            linkcolor=blue,
            citecolor=blue,
            bookmarks=true,
            bookmarksopen=true,
            bookmarksnumbered=true]{hyperref}

\begin{document}
%\begin{CJK}{GBK}{song}
\title{The $X(3960)$, $X_0(4140)$, and other compact $cs\bar{c}\bar{s}$ states}
\author{Shi-Yuan Li$^1$}
\author{Yan-Rui Liu$^1$}\email{yrliu@sdu.edu.cn}
\author{Zi-Long Man$^1$}\email{manzilong@mail.sdu.edu.cn}
\author{Zong-Guo Si$^1$}
\author{Jing Wu$^2$}\email{wujing18@sdjzu.edu.cn}

\affiliation{$^1$School of Physics, Shandong University, Jinan, Shandong 250100, China\\
$^2$School of Science, Shandong Jianzhu University, Jinan 250101, China
}

\date{\today}

\begin{abstract}
We study the spectrum and rearrangement decays of S-wave $cs\bar{c}\bar{s}$ tetraquark states in a simplified quark model. The masses and widths are estimated by assuming that the $X(4140)$ is the lower $1^{++}$ $cs\bar{c}\bar{s}$ tetraquark. Comparing our results with experimental measurements, we find that the recently observed $X(3960)$ by LHCb can be assigned as the lowest $0^{++}$ $cs\bar{c}\bar{s}$ tetraquark state and the $X_0(4140)$ could be the second lowest $0^{++}$ $cs\bar{c}\bar{s}$ tetraquark. Predictions of ratios between partial widths for the involved tetraquarks are given. We call for searches for more $cs\bar{c}\bar{s}$ tetraquarks with $J^{PC}=1^{+-}$, $0^{++}$, and $2^{++}$.
\end{abstract}

%\pacs{xxx}

%\end{CJK}
\maketitle

%%%%%%%%%%%%%%%%%%%%%%%%%%%%%%%%%%%%%%%%%%%
\section{Introduction}\label{sec1}
%%%%%%%%%%%%%%%%%%%%%%%%%%%%%%%%%%%%%%%%%%%

Recently, a near-threshold peaking structure $X(3960)$ was observed in the $D_s^+D_s^-$ invariant mass distribution in the decay $B^+\to D_s^+D_s^-K^+$ by the LHCb Collaboration \cite{LHCb:2022vsv}. The measured mass and width are $M=3956\pm5\pm10$ MeV and $\Gamma=43\pm13\pm8$ MeV, respectively. Its quantum numbers $J^{PC}=0^{++}$ are favored over $1^{--}$ and $2^{++}$. The LHCb analysis indicates that this structure is an exotic candidate consisting of $cs\bar{c}\bar{s}$ constituent. In the same process, the LHCb also reported evidence of another structure $X_0(4140)$ with mass $4133\pm6\pm6$ MeV, width $\Gamma=67\pm17\pm7$ MeV, and quantum numbers $J^{PC}=0^{++}$.

Before the observation of $X(3960)$, exotic states having $cs\bar{c}\bar{s}$ quark component have been reported in the $J/\psi\phi$ invariant mass distributions by various experiment collaborations. In 2008, the CDF Collaboration firstly announced the evidence of a structure with mass $M=4143.0\pm2.9\pm1.2$ MeV and width $\Gamma= 11.7^{+8.3}_{-5.0}\pm3.7$ MeV in the decay $B^{+}\to J/\psi\phi K^{+}$, which was named as $X(4140)$ \cite{CDF:2009jgo}. Later, the CMS Collaboration \cite{CMS:2013jru} and D0 Collaboration \cite{D0:2013jvp} confirmed the $X(4140)$ in the same process, but the Belle Collaboration \cite{Cheng-Ping:2009sgk}, BABAR Collaboration \cite{BaBar:2014wwp}, and LHCb Collaboration \cite{LHCb:2012wyi} did not obtain positive results for this state. From an analysis for the process $\gamma\gamma\to J/\psi\phi$ \cite{Belle:2009rkh}, the Belle got a narrow structure $X(4350)$ which has a mass $M=4350.6^{+4.6}_{-5.1}\pm0.7$ MeV and width $\Gamma=13^{+18}_{-9}\pm4$ MeV. In the decay $B\to J/\psi\phi K$, the CDF observed the evidence of a second $J/\psi\phi$ structure with mass $M=4274^{+8.4}_{-6.7}\pm1.9$ MeV and width $\Gamma=32.3^{+21.9}_{-15.3}\pm7.6$ MeV \cite{CDF:2011pep} while CMS reported the evidence with mass $M=4313.8\pm5.3\pm7.3$ MeV and width $\Gamma=38^{+30}_{-15}\pm15$ MeV \cite{CMS:2013jru,Chistov:2022rht}.  With more collected data, the LHCb searched for $J/\psi\phi$ structures in the decay $B^+\to J/\psi\phi K^+$ again in Ref. \cite{LHCb:2016axx}. The collaboration confirmed the $X(4140)$ with a broader width whose quantum numbers are determined to be $J^{PC}=1^{++}$, established the existence of the $X(4274)$ with $J^{PC}=1^{++}$, and observed two higher resonances $X(4500)$ and $X(4700)$ with $J^{PC}=0^{++}$. In an improved analysis of $B^+\to J/\psi\phi K^+$ \cite{LHCb:2021uow}, the LHCb observed two more states $X(4685)$ and $X(4630)$. The quantum numbers for the former state are $J^P=1^+$ while the preferred $J^P$ for the latter are $1^-$.

Since the observation of $X_1(4140)$\footnote{To distinguish between the two states around 4140 MeV, we will use $X_1(4140)$ to denote the $X(4140)$ with $J^{PC}=1^{++}$ in the following discussions.}, various theoretical explanations such as compact $cs\bar{c}\bar{s}$ tetraquarks and $D_s^{*+}D_s^{*-}$ molecules have been proposed to understand the above exotic resonances in different methods \cite{Liu:2009ei,Mahajan:2009pj,Branz:2009yt,Albuquerque:2009ak,Ding:2009vd,Danilkin:2009hr,Liu:2010hf,Stancu:2009ka,Wang:2015pea,Wang:2016gxp,Chen:2016oma,Maiani:2016wlq,Zhu:2016arf,Lu:2016cwr,Wu:2016gas,Yang:2019dxd,Branz:2010rj,Hao:2019fjg,Deng:2017xlb,Deng:2019dbg,Shi:2021jyr,Yang:2021sue,Liu:2021xje,Wang:2021ghk,Turkan:2021ome,Nakamura:2021bvs,Ge:2021sdq,Ferretti:2021xjl,Agaev:2022iha,Bayar:2022dqa,Ji:2022uie,Ji:2022vdj,Xin:2022bzt,Chen:2022dad,Agaev:2022pis,Mutuk:2022ckn,Xie:2022ilz,Guo:2022zbc,Yang:2022zxe,Badalian:2023qyi,Chen:2023eix,Agaev:2023gti,Guo:2022crh}. Because of their high masses, the $X(4500)$, $X(4700)$, $X(4685)$, and $X(4630)$ may be interpreted as the orbitally or radially excited tetraquark or molecular states \cite{Chen:2016oma,Maiani:2016wlq,Lu:2016cwr,Deng:2017xlb,Liu:2021xje,Yang:2021sue,Wang:2021ghk,Turkan:2021ome,Agaev:2022iha}. For the $X_1(4140)$ and $X(4274)$, the measured quantum numbers $J^{PC}=1^{++}$ \cite{LHCb:2021uow} do not support their $D_s^{*+}D_s^{*-}$ molecule interpretation. For the newly observed $X(3960)$, the authors of Ref. \cite{Bayar:2022dqa,Ji:2022uie,Ji:2022vdj} interpret it as a hadronic molecule in the coupled $D\bar{D}-D_s^+D_s^-$ system. The calculations in the QCD sum rule method \cite{Xin:2022bzt,Mutuk:2022ckn} and in the one-boson-exchange model \cite{Chen:2022dad} also favor the molecule interpretation. Another calculation with QCD two-point sum rules \cite{Agaev:2022pis} leads to the assignment that the $X(3960)$ is a scalar diquark-antidiquark state. From investigations on the $D_s^+D_s^-$ invariant mass spectrum and the ratio $\Gamma(X\to D^+D^-)/\Gamma(D_s^+D_s^-)$, the authors of Ref. \cite{Guo:2022zbc} propose that the $X(3960)$ is induced by the $\chi_{c0}(2P)$ charmonium below the $D_s^+D_s^-$ threshold. A combined analysis in Ref. \cite{Chen:2023eix} by assuming that the $X(3930)$, $X(3960)$, and $X(3915)$ are the same hadron indicates that this state probably has a $c\bar{c}$ core strongly renormalized by the $D_s^+D_s^-$ coupling. In Ref. \cite{Badalian:2023qyi}, the $X(3960)$ and $X_0(4140)$ are interpreted as four-quark $cs\bar{c}\bar{s}$ states while the conclusion that the $X_0(4140)$ is a $D_s^+D_s^-$ molecule is drawn in Ref. \cite{Agaev:2023gti}. The investigation in an improved chromomagnetic interaction model indicates that both $X(3960)$ and $X_0(4140)$ may be interpreted as $0^{++}$ $cs\bar{c}\bar{s}$ tetraquark states \cite{Guo:2022crh}.

To understand the near-threshold structures $X(3960)$ and $X_0(4140)$ and consider other possible tetraquark states, it is worthwhile to further study the $cs\bar{c}\bar{s}$ tetraquark states systematically. In our previous work \cite{Wu:2016gas}, we investigated the spectrum of $cs\bar{c}\bar{s}$ states with a color-magnetic interaction (CMI) model. In the spectrum, there are two $J^{PC}=1^{++}$ and four $J^{PC}=0^{++}$ states. Our results indicate that the $X_1(4140)$ and $X(4274)$ can be interpreted as these two axial-vector tetraquarks while the $X(4350)$ can be assigned as the highest scalar tetraquark. To gain more information about multiquark states, their decay properties deserve to be investigated. We tried to describe the rearrangement decays of hidden-charm pentaquark states \cite{Cheng:2019obk,Li:2023aui} and tetraquark states with different flavors \cite{Cheng:2020nho} in a simple scheme, where the constant Hamiltonian $H_{decay}={\cal C}$ was adopted. In the present work, we update our previous study on the compact $cs\bar{c}\bar{s}$ states by including the decay properties. We will identify the $X_1(4140)$ as the low $1^{++}$ or the $X(4274)$ as the high $1^{++}$ $cs\bar{c}\bar{s}$ tetraquark and use their masses and widths as inputs to discuss the massed and widths of $cs\bar{c}\bar{s}$ states.

This paper is organized as follows. In Sec. \ref{sec2}, we present the CMI Hamiltonian, wave functions, and method to consider rearrangement decays. In Sec. \ref{sec3}, we give the model parameters and numerical results. The last section is for some discussions and a summary.

%%%%%%%%%%%%%%%%%%%%%%%%%%%%%%%%%%%%%%%%%%%
\section{Formalism }\label{sec2}
%%%%%%%%%%%%%%%%%%%%%%%%%%%%%%%%%%%%%%%%%%%

\subsection{Model Hamiltonian and wave functions}

The effective Hamiltonian in the CMI model to study the mass spectrum of the $cs\bar{c}\bar{s}$ tetraquark states reads
\begin{eqnarray}\label{hamiltonian}
H=\sum_i m_i+H_{\mathrm{CMI}}=\sum_i m_i-\sum_{i<j}C_{ij} \lambda_i\cdot\lambda_j\sigma_i\cdot\sigma_j
\end{eqnarray}
where $m_i$ is effective mass of the $i$th quark and $C_{ij}$ denotes the effective coupling constant between the $i$th quark and $j$th quark. $\lambda_i$ and $\sigma_i$ stand for the SU(3) Gell-Mann matrices and SU(2) Pauli matrices for the $i$th quark, respectively. For an antiquark, $\lambda_i\to -\lambda_i^*$. The mass splittings between different $cs\bar{c}\bar{s}$ states are mainly induced by the term $H_{\mathrm{CMI}}$. The CMI model is a simplified quark model where $m_i$ contains the constituent quark mass and contributions from the kinetic energy, color-Coulomb, and linear confinement terms.

After getting the eigenvalue $E_{\mathrm{CMI}}$ of $H_{\mathrm{CMI}}$, one obtains the mass formula for a tetraquark state,
\begin{eqnarray}\label{mref}
M =\sum_i m_i+E_{\mathrm{CMI}}.
\end{eqnarray}
From numerical results for various systems \cite{Liu:2019zoy,Wu:2016gas,Wu:2017weo,Wu:2016vtq,Wu:2018xdi,Li:2018vhp,Cheng:2020nho}, one finds that the estimated masses of either conventional hadrons or multiquark states using this equation are higher than measured values. The main reason is that the values of effective mass $m_i$ in different systems should actually be different. It indicates that the effective attraction between quark components is not appropriately taken into consideration. These overestimated values can be regarded as upper limits for the multiquark masses on the theoretical side. To reduce the uncertainties in the CMI model, one may include the color-electric term explicitly. Here, we use the modified formula to explore the mass splittings between different $cs\bar{c}\bar{s}$ states by introducing a reference system,
\begin{eqnarray}
M =[M_{ref}-(E_{\mathrm{CMI}})_{ref}]+E_{\mathrm{CMI}},
\end{eqnarray}
where $M_{ref}$ and $(E_{\mathrm{CMI}})_{ref}$ denote the measured mass and calculated CMI eigenvalue for the reference system, respectively. One may adopt a meson-meson state as the reference whose quark content is the same as the considered tetraquark, but the reasonable choice is hard to make. In Ref. \cite{Wu:2016gas}, we estimated the massed of $cs\bar{c}\bar{s}$ tetraquarks using the $J/\psi\phi$ and $D^{+}_sD^{*-}_s$ thresholds and find that the resulting $1^{++}$ tetraquark masses in both cases are lower than the experimental measurement. The results with $J/\psi\phi$ are about 100 MeV lower than those with $D^{+}_sD^{*-}_s$. So we may treat these underestimated values as lower limits for theoretical $cs\bar{c}\bar{s}$ masses. On the other hand, if one chooses the $X_1(4140)$ as the reference state by assigning it to be the lower $1^{++}$ $cs\bar{c}\bar{s}$ tetraquark, the assignment for $X(4274)$ as the higher $1^{++}$ $cs\bar{c}\bar{s}$ is acceptable. Along this line, we have extended studies using the $X_1(4140)$ as input to other tetraquark systems \cite{Wu:2018xdi,Cheng:2020nho}. In the present work, we still follow this idea and update previous study \cite{Wu:2016gas} by including the decay information. For this purpose, we need to know the spin$\otimes$color wave functions.

For a $cs\bar{c}\bar{s}$ tetraquark, the total wave function is not constrained by the Pauli principle, but one must consider its C-parity since it is a truly neutral state. We have given the spin$\otimes$color wave functions in Ref. \cite{Wu:2016gas} in the diquark-antidiquark base. For convenience, we present here the definitions again. With the notation $[(cs)_{\mathrm{color}}^{\mathrm{spin}}(\bar{c}\bar{s})_{\mathrm{color}}^{\mathrm{spin}}]^{\mathrm{spin}}$, they are
\begin{eqnarray}
{J^{PC}=2^{++}}: \quad &\phi_1\chi_1=[(cs)_{6}^{1}(\bar{c}\bar{s})_{\bar{6}}^{1}]^{2},\quad \phi_2\chi_1=[(cs)_{\bar{3}}^{1}(\bar{c}\bar{s})_{3}^{1}]^{2};\notag\\
{J^{PC}=0^{++}}:\quad &\phi_1\chi_3=[(cs)_{6}^{1}(\bar{c}\bar{s})_{\bar{6}}^{1}]^{0},\quad \phi_2\chi_3=[(cs)_{\bar{3}}^{1}(\bar{c}\bar{s})_{3}^{1}]^{0}, 
&\phi_1\chi_6=[(cs)_{6}^{0}(\bar{c}\bar{s})_{\bar{6}}^{0}]^{0},\quad\phi_2\chi_6=[(cs)_{\bar{3}}^{0}(\bar{c}\bar{s})_{3}^{0}]^{0};
\end{eqnarray}
\begin{eqnarray}
{J^{PC}=1^{++}}:\quad &\phi_1\chi_+=\frac{1}{\sqrt{2}}\Big([(cs)_{6}^{1}(\bar{c}\bar{s})_{\bar{6}}^{0}]^{1}+[(cs)_{6}^{0}(\bar{c}\bar{s})_{\bar{6}}^{1}]^{1}\Big),\quad \phi_2\chi_+=\frac{1}{\sqrt{2}}\Big([(cs)_{\bar{3}}^{1}(\bar{c}\bar{s})_{3}^{0}]^{1}+[(cs)_{\bar{3}}^{0}(\bar{c}\bar{s})_{3}^{1}]^{1}\Big);\notag\\
{J^{PC}=1^{+-}}:\quad &\phi_1\chi_-=\frac{1}{\sqrt{2}}\Big([(cs)_{6}^{1}(\bar{c}\bar{s})_{\bar{6}}^{0}]^{1}-[(cs)_{6}^{0}(\bar{c}\bar{s})_{\bar{6}}^{1}]^{1}\Big),\quad \phi_2\chi_-=\frac{1}{\sqrt{2}}\Big([(cs)_{\bar{3}}^{1}(\bar{c}\bar{s})_{3}^{0}]^{1}-[(cs)_{\bar{3}}^{0}(\bar{c}\bar{s})_{3}^{1}]^{1}\Big),\notag\\
 &\phi_1\chi_2=[(cs)_{6}^{1}(\bar{c}\bar{s})_{\bar{6}}^{1}]^{1},\quad
 \phi_2\chi_2=[(cs)_{\bar{3}}^{1}(\bar{c}\bar{s})_{3}^{1}]^{1}.
\end{eqnarray}
We do not give the explicit CMI matrices here. One can find in Ref. \cite{Wu:2016gas} the matrices with the bases $(\phi_1\chi_1, \phi_2\chi_1)^{T}$, $(\phi_1\chi_+, \phi_2\chi_+)^{T}$,  $(\phi_1\chi_3, \phi_2\chi_3, \phi_1\chi_6, \phi_2\phi_6)^{T}$, and $(\phi_1\chi_2, \phi_2\chi_2, \phi_1\chi_-, \phi_2\chi_-)^{T}$ for the $J^{PC}=2^{++}$, $1^{++}$, $0^{++}$, and $1^{+-}$ cases, respectively.

\subsection{Effective interactions and rearrangement decays}

To reflect the effective CMI between quark components in multiquark states, various $K$ factors were introduced in Ref. \cite{Li:2018vhp}. Later in Ref. \cite{Wu:2018xdi}, we calculated the $K$ factors for the $cs\bar{c}\bar{s}$ states. With them, we argued that the highest $2^{++}$, the highest $1^{++}$, and the second highest $0^{++}$ states are probably more stable than other partners. Whether the argument is sound or not will be checked in the next section. The $K$ factor between the $i$th and $j$th quark components is given by
\begin{eqnarray}
K_{ij}=\lim_{\Delta C_{ij}\to 0}\frac{\Delta E_{\mathrm CMI}}{\Delta C_{ij}}\to \frac{\partial E_{\mathrm CMI}}{\partial C_{ij}},
\end{eqnarray}
where $\Delta C_{ij}$ is the variation of an effective coupling constant and $\Delta E_{\mathrm CMI}$ is the corresponding variation of a multiquark mass. Now, the mass of a tetraquark state can be rewritten as
\begin{eqnarray}
M=[M_{ref}-(E_{\mathrm CMI})_{ref}]+\sum_{i<j}K_{ij}C_{ij}
\end{eqnarray}
The sign of $K_{ij}$ reflects whether the effective CMI \cite{Wu:2016vtq} between the $i$th and $j$th quark components is attractive ($K_{ij}<0$) or repulsive ($K_{ij}>0$).

The strong decays of conventional hadrons involve the creation of at least one quark-antiquark pair at the quark level. One needs to choose a quark creation mechanism in the calculation. The $^3P_0$ model is usually adopted in studying the two-body strong decays where a unique coupling constant is used. For the strong decays of compact tetraquark states, the two-body decay patterns should be the dominant ones but they do not involve quark creations. In this work, we use a simple method to calculate the rearrangement decay widths of $cs\bar{c}\bar{s}$ states, where the quark-level Hamitonian for decay is taken as a constant $H_{decay}=\mathcal {C}$. It means that the four quark components in different tetraquarks scatter to meson-meson states freely with equal coupling strength. This method has been applied to deal with decays of pentaquarks states \cite{Cheng:2019obk,Li:2023aui} and tetraquark states with four different flavors \cite{Cheng:2020nho}. In principle, gluon exchanges would induce corrections to this simple model, but additional parameters are also needed. At present, we assume that the rearrangement decays for all the $cs\bar{c}\bar{s}$ tetraquark states can be described by this single constant ${\cal C}$. Because there is only one parameter, the partial width ratio is a good quantity to test the model. If more experimental decay data are available, modification of the decay Hamiltonian may be considered.

In the adopted model, the width for a rearrangement decay channel is
\begin{gather}
\Gamma=\frac{\sqrt{(M^2-(m_1+m_2)^2)(M^2-(m_1-m_2)^2)}}{16\pi M^3}|\mathcal{M}|^2,
\end{gather}
where $M$, $m_1$, and $m_2$ are the masses of initial tetraquark and two final mesons, respectively. The decay amplitude $\mathcal{M}=\langle initial|H_{decay}|final \rangle$ is given by
\begin{eqnarray}
\mathcal{M}=\mathcal{C}\sum_{ij}\alpha_i\beta_j
\end{eqnarray}
where $\alpha_i$'s illustrated in the following Eq. \eqref{initialstates} are coefficients of the initial wave function in the bases given in last subsection and $\beta_j$'s denote coefficients of the final meson-meson wave function in the same bases. For the initial states, their spin-color wave functions have the forms
\begin{eqnarray}\label{initialstates}
\Psi(2^{++}) &=& \alpha_1\phi_1\chi_1+ \alpha_2\phi_2\chi_1,\nonumber\\
\Psi(1^{++}) &=& \alpha_1\phi_1\chi_+ +\alpha_2\phi_2\chi_+,\nonumber\\
\Psi(0^{++}) &=& \alpha_1\phi_1\chi_3+ \alpha_2\phi_2\chi_3+\alpha_3\phi_1\chi_6+\alpha_4\phi_2\chi_6,\nonumber\\
\Psi(1^{+-}) &=& \alpha_1\phi_1\chi_2+\alpha_2\phi_2\chi_2+\alpha_3\phi_1\chi_-+\alpha_4\phi_2\chi_-,
\end{eqnarray}
where the normalization condition $\sum_{i=1}|\alpha_i|^2=1$ is always satisfied. One gets the values of $\alpha_i$'s from the eigenvector of the corresponding tetraquark CMI matrix. We show them explicitly in the following table \ref{mass}. There are two types of rearrangement decays $Q_1q_2\bar{Q}_3\bar{q}_4\to(Q_1\bar{Q}_3)_{1c}+(q_2\bar{q}_4)_{1c}$ and $Q_1q_2\bar{Q}_3\bar{q}_4\to(Q_1\bar{q}_4)_{1c}+(q_2\bar{Q}_3)_{1c}$. One gets the $\beta_j$'s by recoupling the final meson-meson states into the forms similar to those in Eq. \eqref{initialstates}. Since we project out the initial bases from the final wave functions, one needs two recoupling formulas in the color space,
\begin{eqnarray}
\begin{array}{l}
(Q_1\bar{Q}_3)_{1c}(q_2\bar{q}_4)_{1c}=-\frac{1}{\sqrt{3}}\phi_1+\sqrt{\frac{2}{3}}\phi_2,\\
(Q_1\bar{q}_4)_{1c}(q_2\bar{Q}_3)_{1c}=\frac{1}{\sqrt{3}}\phi_1+\sqrt{\frac{2}{3}}\phi_2.
\end{array}
\end{eqnarray}
In the spin space, similar formulas are easy to get by calculating the $9j$ symbols. Then the explicit values of $\beta_j$'s are obtained with these two-space coefficients.

\setlength{\tabcolsep}{0.5mm}

%%%%%%%%%%%%%%%%%%%%%%%%%%%%%%%%%%%%%%%%%%%
\section{Model parameters and numerical results}\label{sec3}
%%%%%%%%%%%%%%%%%%%%%%%%%%%%%%%%%%%%%%%%%%%

\begin{table}[!htb]
	\caption{Chromomagnetic interactions for relevant hadrons and the obtained coupling parameters in units of MeV.}\setlength{\tabcolsep}{1.3mm}\label{effectiveparameters}
	\centering
	\begin{tabular}{ccccc}
		\hline\hline
  	Hadron      &$E_{CMI}$          &Hadron   &$E_{CMI}$&$C_{ij}$ \\\hline
  	  $\Xi_c^{\prime}$ &$\frac83C_{ns}-\frac{16}{3}C_{cn}-\frac{16}{3}C_{cs}$&$\Xi_c^*$&$\frac83C_{ns}+\frac{8}{3}C_{cn}+\frac{8}{3}C_{cs}$&$C_{cs}=4.49\pm 0.08$\\
   $D_s$&$-16C_{c\bar{s}}$&$D^*_s$&$\frac{16}{3}C_{c\bar{s}}$&$C_{c\bar{s}}=6.75\pm 0.02$ \\
  	  $\eta_c$&$-16C_{c\bar{c}}$&$J/\psi$&$\frac{16}{3}C_{c\bar{c}}$&$C_{c\bar{c}}=5.30\pm 0.02$\\
  	  $2\Omega+\Delta-(2\Xi^*+\Xi)$&$8C_{ss}+8C_{nn}$&$(\Delta-N)/2$&$8C_{nn}$&$C_{ss}=6.46\pm 0.11$\\
  	  \hline \hline
	\end{tabular}\\
\end{table}

The coupling parameters $C_{cs}$, $C_{c\bar{s}}$, $C_{c\bar{c}}$, and $C_{ss}$ that we adopt in estimating the $cs\bar{c}\bar{s}$ masses are extracted from the measured masses of the conventional ground hadrons \cite{ParticleDataGroup:2022pth} by using their mass formulas in the CMI model. The relevant hadrons, their $E_{CMI}$'s, and the determined coupling parameters are collected in table \ref{effectiveparameters}. One gets other coupling parameters similarly \cite{Wu:2016gas,Wu:2018xdi}. In table \ref{effectiveparameters}, the errors of $C_{ij}$'s are also presented. Because the systematic error of the CMI model cannot be estimated and it might be larger than the measurement error, we do not consider errors in the following numerical estimations. We just take $C_{cs}=4.5$ MeV, $C_{c\bar{s}}=6.8$ MeV, $C_{c\bar{c}}=5.3$ MeV, and $C_{ss}=6.5$ MeV. The adopted coupling parameters $C_{cc}$ and $C_{s\bar{s}}$ are obtained with the approximation $\frac{C_{cc}}{C_{c\bar{c}}}=\frac{C_{ss}}{C_{s\bar{s}}}=\frac{C_{nn}}{C_{n\bar{n}}}\approx\frac23$. To estimate the upper limit masses, we use the effective quark masses $m_s=(M_\Omega-8C_{ss})/3=542.4$ MeV and $m_c=(3M_{\Sigma_c^*}-2M_\Delta-16C_{nc}+8C_{nn})/3=1724.1$ MeV where $n$ indicates $u$ or $d$ quark. The extraction details can be found in Refs. \cite{Wu:2016gas,Wu:2018xdi}. In the estimation of lower limits for the masses with the $J/\psi\phi$ threshold, one does not need $m_s$ and $m_c$.

When one estimates the masses of other $cs\bar{c}\bar{s}$ states using the $X_1(4140)$, the input mass needs to be determined. In Ref. \cite{LHCb:2016axx}, the mass and width of $X_1(4140)$ determined by LHCb are $4146.5\pm 4.5^{+4.6}_{-2.8}$ MeV and $83\pm 21^{+21}_{-14}$ MeV, respectively. In Ref. \cite{LHCb:2021uow}, the values are updated to $4118\pm11^{+19}_{-36}$ MeV and $162\pm21^{+24}_{-49}$ MeV, respectively. In the particle data book \cite{ParticleDataGroup:2022pth}, the values averaged from different measurements are $4146.5\pm 3.0$ MeV and $19^{+7}_{-5}$ MeV, respectively. Although the experimental masses are all around 4140 MeV, the deviation in width is significant. For the other $1^{++}$ state, $X(4274)$, the deviation in width between different collaborations is insignificant \cite{ParticleDataGroup:2022pth}. In Ref. \cite{Cheng:2020nho}, we chose the LHCb results in Ref. \cite{LHCb:2016axx} as inputs from the consistency consideration for widths between $X_1(4140)$ and $X(4274)$. The necessary condition for our purpose is that $\Gamma(X_1(4140))$ and $\Gamma(X(4274))$ are comparable. Here, we still follow Ref. \cite{Cheng:2020nho} and use data determined in Ref. \cite{LHCb:2016axx}. The cases with other choices will also be discussed. In addition to using the $X_1(4140)$ as reference state, we will discuss the case using the $X(4274)$ as input, too.

The rearrangement decay channels for a $1^{++}$ $cs\bar{c}\bar{s}$ state are $J/\psi\phi$ and $\frac{1}{\sqrt2}(D_s^{*+}D_s^--D_s^+D_s^{*-})$ where the convention for relative phase \cite{Liu:2013rxa} is determined with $D_s^{(*)+}=c\bar{s}$ and $D_s^{(*)-}=s\bar{c}$. Assuming that the total decay width of a tetraquark is equal to the sum of partial widths ($\Gamma_{sum}$) for rearrangement decay channels, one extracts $\mathcal{C}=72822$ MeV from the LHCb data \cite{LHCb:2016axx}.

The final states for the decay of $cs\bar{c}\bar{s}$ tetraquarks involve conventional mesons containing the $s\bar{s}$ component. In the quark model, the quark content of the vector meson $\phi$ is almost $s\bar{s}$, but that of the pseudoscalar mesons $\eta$ and $\eta^\prime$ not. They are superpositions of SU(3) singlet state $\eta_1$ and octet state $\eta_8$,
\begin{eqnarray}
|\eta\rangle&=&cos(\theta)|\eta_8\rangle-sin(\theta)|\eta_1\rangle ,\nonumber\\
|\eta^\prime\rangle&=&sin(\theta)|\eta_8\rangle+cos(\theta)|\eta_1\rangle,
\end{eqnarray}
where $\theta$ is the mixing angle. We employ the value $\theta=-11.3^{\circ}$ \cite{ParticleDataGroup:2022pth} in our calculation.

\begin{figure}[htbp]\centering
	\includegraphics[width=260pt]{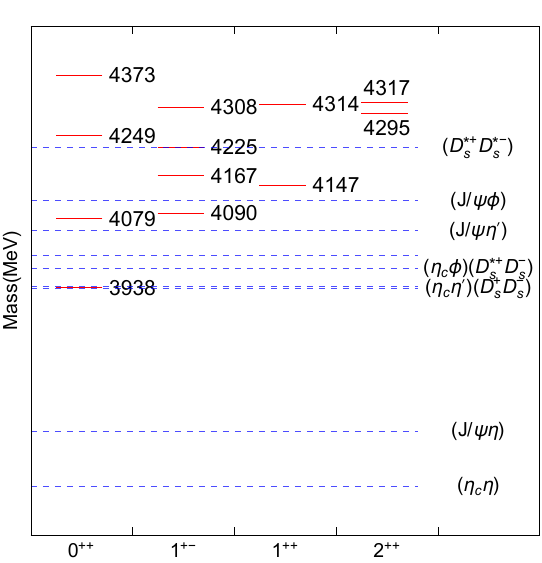}
	\caption{Relative positions for the $cs\bar{c}\bar{s}$ tetraquark states. The red solid and blue dashed lines correspond to estimated masses (with $X_1(4140)$) and related meson-meson thresholds, respectively.}\label{cscs-picture}
\end{figure}

With the above parameters, we obtain the numerical results for ground $cs\bar{c}\bar{s}$ states. The mass results are collected in Table \ref{mass}. Comparing with Ref. \cite{Wu:2016gas}, one finds some differences in number which result mainly from the variation of coupling parameters. We show the relative positions for the $cs\bar{c}\bar{s}$ states using the input $X_1(4140)$ in Fig. \ref{cscs-picture}. The related meson-meson thresholds are also displayed. The results for the rearrangement decays are given in Table \ref{decay1}.

\setlength{\tabcolsep}{2mm}\begin{table}[!h]
\caption{Numerical results for the masses of $cs\bar{c}\bar{s}$ states in units of MeV. The bases for $\langle H_{CMI}\rangle$ in the $2^{++}$, $1^{++}$, $0^{++}$, and $1^{+-}$ cases are $(\phi_1\chi_1,\phi_2\chi_1)^T$, $(\phi_1\chi_+,\phi_2\chi_+)^T$, $(\phi_1\chi_3,\phi_2\chi_3,\phi_1\chi_6,\phi_2\chi_6)^T$, and $(\phi_1\chi_2,\phi_2\chi_2,\phi_1\chi_-,\phi_2\chi_-)^T$, respectively. The masses obtained with $X_1(4140)$ are given in the fifth column. The lower limits and upper limits for the masses are listed in the sixth and seventh columns, respectively.}\scriptsize\label{mass}
\begin{tabular}{c|ccccccc}\hline
\hline
$J^{PC}$ & $\langle H_{CMI} \rangle$ &Eigenvalue &Eigenvector &$\mathrm{Mass}$&Lower limits&Upper limits\\\hline
$2^{++}$ &$\left(\begin{array}{cc}62.8&-4.2\\-4.2&83.4\end{array}\right)$&$\left(\begin{array}{c}84.2\\62.0\end{array}\right)$&$\left(\begin{array}{cc}\{-0.19,0.98\}\\\{-0.98,-0.19\}\end{array}\right)$&$\left(\begin{array}{c}4316.9\\4294.6\end{array}\right)$&$\left(\begin{array}{c}4120.1\\4097.8\end{array}\right)$&$\left(\begin{array}{c}4617.2\\4595.0\end{array}\right)$\\
$1^{++}$ &$\left(\begin{array}{cc}-22.5&-81.2\\-81.2&17.3\end{array}\right)$&$\left(\begin{array}{c}80.9\\-86.2\end{array}\right)$&$\left(\begin{array}{cc}\{0.62,-0.79\}\\\{-0.79,-0.62\}\end{array}\right)$&$\left(\begin{array}{c}4313.6\\4146.5\end{array}\right)$&$\left(\begin{array}{c}4116.8\\3949.6\end{array}\right)$&$\left(\begin{array}{c}4613.9\\4446.8\end{array}\right)$\\
$0^{++}$ &$\left(\begin{array}{cccc}-52.0&8.5&-3.5&140.6\\8.5&-203.6&140.6&-8.7\\-3.5&140.6&-73.6&0\\140.6&-8.7&0&36.8\end{array}\right)$&$\left(\begin{array}{c}139.9\\16.3\\-154.2\\-294.5\end{array}\right)$&$\left(\begin{array}{cccc}\{0.59,-0.01,-0.02,0.81\}\\\{-0.03,-0.54,-0.84,-0.00\}\\\{-0.80,-0.06,0.06,0.59\}\\\{0.07,-0.84,0.54,-0.05\}\end{array}\right)$&$\left(\begin{array}{c}4372.6\\4249.0\\4078.5\\3938.2\end{array}\right)$&$\left(\begin{array}{c}4175.7\\4052.2\\3881.6\\3741.4\end{array}\right)$&$\left(\begin{array}{c}4672.9\\4549.3\\4378.8\\4238.5\end{array}\right)$\\
$1^{+-}$ &$\left(\begin{array}{cccc}-13.7&4.2&-12.0&25.5\\4.2&-107.9&25.5&-30.0\\-12.0&25.5&-26.5&81.2\\25.5&-30.0&81.2&7.3\end{array}\right)$&$\left(\begin{array}{c}75.3\\-7.5\\-65.8\\-142.9\end{array}\right)$&$\left(\begin{array}{cccc}\{-0.14,0.04,-0.60,-0.79\}\\\{0.94,-0.07,-0.33,0.08\}\\\{-0.28,-0.66,-0.54,0.43\}\\\{-0.16,0.74,-0.48,0.44\}\end{array}\right)$&$\left(\begin{array}{c}4308.0\\4225.1\\4166.9\\4089.8\end{array}\right)$&$\left(\begin{array}{c}4111.2\\4028.3\\3970.0\\3892.9\end{array}\right)$&$\left(\begin{array}{c}4608.3\\4525.5\\4467.2\\4390.1\end{array}\right)$\\
\hline\hline
\end{tabular}
\end{table}

\setlength{\tabcolsep}{1mm}\begin{table}[htbp]
	\caption{Rearrangement decays for the $cs\bar{c}\bar{s}$ states by assigning the $X_1(4140)$ as the lighter $1^{++}$ $cs\bar{c}\bar{s}$ tetraquark. The numbers in the parentheses are ($100|{\cal M}|^{2}/\mathcal{C}^{2},\Gamma$) where the coupling parameter $\mathcal{C}$ is extracted from the width of $X_1(4140)$ (83 MeV \cite{LHCb:2016axx}). The partial width $\Gamma$ and total width $\Gamma_{sum}$ are given  in units of MeV.}\scriptsize\label{decay1}
	\begin{tabular}{c|c|ccccc|c}\hline\hline
		$J^{PC}$&Mass&\multicolumn{5}{c}{Channels}&$\Gamma_{sum}$\\
		\hline
		&&$J/ \psi \phi$&$D_s^{*+}D_s^{*-}$&&&\\
		$2^{++}$&$\left[\begin{array}{c}4316.9\\4294.6\end{array}\right]$&$\left[\begin{array}{c}(83.4, 53.8)\\(16.6, 10.2)\end{array}\right]$&$\left[\begin{array}{c}(47.5, 23.9)\\(52.5, 23.2)\end{array}\right]$&&&&$\left[\begin{array}{c}77.7\\33.4\end{array}\right]$\\
		&&$J/ \psi \phi$&$( D_s^{*+}D_s^- - D_s^+D_s^{*-} ) /\sqrt{2}$&&&\\
		$1^{++}$&$\left[\begin{array}{c}4313.6\\4146.5\end{array}\right]$&$\left[\begin{array}{c}(99.8, 63.9)\\(0.2, 0.1)\end{array}\right]$&$\left[\begin{array}{c}(8.2, 13.0)\\(91.8, 82.9)\end{array}\right]$&&&&$\left[\begin{array}{c}76.9\\83.0\end{array}\right]$\\
		&&$J/ \psi \phi$&$\eta_c \eta^{\prime}$&$\eta_c \eta$&$D_s^{*+}D_s^{*-}$&$ D_s^+D_s^-$\\
		$0^{++}$&$\left[\begin{array}{c}4372.6\\4249.0\\4078.5\\3938.2\end{array}\right]$&$\left[\begin{array}{c}(57.1, 40.9)\\(39.5, 21.2)\\(3.1, -)\\(0.3, -)\end{array}\right]$&$\left[\begin{array}{c}(0.0, 0.0)\\(0.8, 0.7)\\(18.0, 10.4)\\(34.0, -)\end{array}\right]$&$\left[\begin{array}{c}(0.0, 0.0)\\(0.7, 0.8)\\(16.0, 16.6)\\(30.3, 28.2)\end{array}\right]$&$\left[\begin{array}{c}(52.8, 32.9)\\(42.7, 11.4)\\(3.8, -)\\(0.8, -)\end{array}\right]$&$\left[\begin{array}{c}(0.1, 0.2)\\(2.3, 2.2)\\(49.2, 33.2)\\(48.4, 3.6)\end{array}\right]$&$\left[\begin{array}{c}74.0\\36.3\\60.2 \\31.8\end{array}\right]$\\
		&&$J/ \psi \eta^{\prime}$&$J/ \psi \eta$&$\eta_c \phi$&$D_s^{*+}D_s^{*-}$&$( D_s^{*+}D_s^- + D_s^+D_s^{*-}  )/\sqrt{2}$\\
		$1^{+-}$&$\left[\begin{array}{c}4308.0\\4225.1\\4166.9\\4089.8\end{array}\right]$&$\left[\begin{array}{c}(0.9, 0.6)\\(3.0, 1.8)\\(2.2, 1.1)\\(46.9, 13.5)\end{array}\right]$&$\left[\begin{array}{c}(0.8, 0.8)\\(2.7, 2.7)\\(1.9, 1.9)\\(41.7, 38.2)\end{array}\right]$&$\left[\begin{array}{c}(8.5, 6.9)\\(36.8, 26.0)\\(54.5, 33.8)\\(0.2, 0.1)\end{array}\right]$&$\left[\begin{array}{c}(97.7, 46.9)\\(1.6, 0.1)\\(0.1, -)\\(0.6, -)\end{array}\right]$&$\left[\begin{array}{c}(0.2, 0.4)\\(23.4, 30.3)\\(49.6, 50.8)\\(26.8, 9.3)\end{array}\right]$&$\left[\begin{array}{c}55.5\\60.9\\87.5 \\61.0\end{array}\right]$\\
		\hline\hline
	\end{tabular}
\end{table}

For the $1^{++}$ $cs\bar{c}\bar{s}$ states, their masses and decays have been discussed in Refs. \cite{Wu:2016gas} and \cite{Cheng:2020nho}, respectively. Although the values are slightly different from those in Tables \ref{mass} and \ref{decay1}, the main conclusion that the $X_1(4140)$ and $X(4274)$ could be consistently interpreted as the two $1^{++}$ $cs\bar{c}\bar{s}$ tetraquarks is remained. The calculated $\Gamma_{sum}=76.3$ MeV for the higher state is slightly larger than the measured width of $51\pm7$ MeV \cite{ParticleDataGroup:2022pth}. It is worth noting that the adopted mass value of $X(4274)$ (4313.6 MeV) is close to the CMS result ($4313.8\pm5.3\pm7.3$ MeV) \cite{CMS:2013jru} but larger than the PDG result ($4286^{+8}_{-9}$ MeV) \cite{ParticleDataGroup:2022pth}. When one adopts the PDG value, the obtained $\Gamma_{sum}$ is 10 MeV smaller and is closer to the measured width. From Table \ref{decay1}, the width ratio between the two channels $J/\psi\phi$ and $D_s\bar{D}_s^*$ for the higher state is $\Gamma(J/\psi\phi)/\Gamma(D_s^*\bar{D}_s)\simeq 4.9$ where $D_s^*\bar{D}_s$ simply means the $C$-even $D_s^{*+}D_s^{-}/D_s^{+}D_s^{*-}$ state, while that for the lower state is $\Gamma(J/\psi\phi)/\Gamma(D_s^*\bar{D}_s)\simeq 10^{-3}$. The hidden-charm decay for the $X(4274)$ is significantly suppressed.

For the two $2^{++}$ $cs\bar{c}\bar{s}$ tetraquarks, their mass gap is 22.3 MeV. The higher state is broader than the lower one. The masses of both states are close to that of $X(4274)$ determined by CMS \cite{CMS:2013jru}. If these two $cs\bar{c}\bar{s}$ mesons do exist, the width ratio for the higher tetraquark between its two rearrangement decay channels is predicted to be
\begin{gather}
\frac{\Gamma(J/\psi\phi)}{\Gamma(D_s^{*+}D_s^{*-})}\simeq 2.3,
\end{gather}
and that for the lower tetraquark would be
\begin{gather}
\frac{\Gamma(J/\psi\phi)}{\Gamma(D_s^{*+}D_s^{*-})}\simeq 0.4.
\end{gather}
These two values are different and the ratio can be used to uncover the nature of a $2^{++}$ exotic state measured in future experiments. The mass gap between the two tetraquarks is smaller than the width of any one. It is also possible that experiments would observe just one state around 4.3 GeV but there are actually two states. The comparison of measurements in future experiments with the above obtained width ratio between $J/\psi\phi$ and $D_s^{*+}D_s^{*-}$ will be helpful to understand possible structures of the observed state(s).

In the $J^{PC}=0^{++}$ case, there are four possible $cs\bar{c}\bar{s}$ tetraquarks. The estimated mass of the highest state ($4372.6$ MeV) is close to the mass of $X(4350)$. This result is consistent with the chiral quark model prediction of Ref. \cite{Yang:2019dxd}. From Table \ref{decay1}, the width of the highest state is about $74$ MeV which is larger than the width of $X(4350)$ ($13^{+18}_{-9}\pm4$ MeV). It should be noted that the experiment value has large uncertainty and we adopt a crude model. Future studies are still needed. At present, we can temporarily assign the $X(4350)$ as the highest $cs\bar{c}\bar{s}$ tetraquark state with quantum numbers $J^{PC}=0^{++}$. In this case, our calculation indicates that its dominant decay channels are $J/\psi\phi$ and $D_s^{*+}D_s^{*-}$, which could be used to test the assignment.

The lowest $0^{++}$ $cs\bar{c}\bar{s}$ tetraquark has mass 3938.2 MeV and width 31.8 MeV. It is a good candidate of the recently reported $X(3960)$. From our results, this scalar tetraquark decays dominantly into the $\eta_c\eta$ channel with the branching fraction Br$[X(3938.2)\to\eta_c\eta=89\%]$. Although the coupling with the channel $D_s^+D_s^-$ is also strong, the suppressed phase space results in a small partial width. With the assignment that the $X(3960)$ is the lowest scalar $cs\bar{c}\bar{s}$ tetraquark state, we predict the decay ratio
\begin{gather}
\frac{\Gamma(\eta_c\eta)}{\Gamma(D_s^+D_s^-)}\simeq 7.8.
\end{gather}
The search for $X(3960)$ in the $\eta_c\eta$ and $D_s^+D_s^-$ channels and the check of this ratio can help us to better understand the nature of this exotic state.

The mass and width of the second lowest $0^{++}$ $cs\bar{c}\bar{s}$ tetraquark are estimated to be $4078.5$ MeV and $60.2$ MeV, respectively, in our model. Its mass is about 55 MeV smaller than the $X_0(4140)$ \cite{LHCb:2022vsv}, but its width is consistent with the $X_0(4140)$. If the $X_0(4140)$ can be interpreted as this $cs\bar{c}\bar{s}$ tetraquark, the ratios between different partial widths are
\begin{gather}
\Gamma(\eta_c\eta^{\prime}):\Gamma(\eta_c\eta):\Gamma(D_s^+D_s^-)\simeq 1:1.6:3.2,
\end{gather}
which can be tested in future experiments.

The second highest $0^{++}$ state has mass $4249.0$ MeV and width $36.3$ MeV. At present, no experimentally observed state can be related to this tetraquark, but its existence is possible. Although the $X(4274)$ has a similar mass and width, the quantum numbers are different. Further search of a $cs\bar{c}\bar{s}$ state around 4250 MeV in the channel $J/\psi\phi$, $\eta_c\eta^{\prime}$, $\eta_c\eta$, $D_s^{*+}D_s^{*-}$, or $D_s^+D_s^-$ is strongly called for.

In the $1^{+-}$ case, there are four $cs\bar{c}\bar{s}$ tetraquark states. From Table \ref{decay1}, the widths of these state are all around 50$\sim$90 MeV. For the lightest state, the coupling with the $\eta_c\phi$ channel is weak and the corresponding partial width is tiny. Then this tetraquark has three dominant rearrangement decay channels. The width ratios between them are
\begin{gather}
\Gamma(J/\psi\eta^{\prime}): \Gamma(J/\psi\eta):\Gamma(D_s^*\bar{D}_s)\simeq 1.5:4.1:1.0,
\end{gather}
where $D_s^*\bar{D}_s$ simply means the $C$-odd $D_s^{*+}D_s^-/D_s^+D_s^{*-}$ state. For the second lowest state, its mass is close to that of $X_1(4140)$. One may choose $J/\psi\eta^{\prime}$, $J/\psi\eta$, $\eta_c\phi$, and $D_s^*\bar{D}_s$ to detect this tetraquark. Its dominant decay modes are $\eta_c\phi$ and $D_s^*\bar{D}_s$. Their ratio $\Gamma(\eta_c\phi):\Gamma(D_s^*\bar{D}_s)\simeq 0.7$ is predicted. For the second highest state, it is around the threshold of $D_s^{*+}D_s^{*-}$ and it has two dominant rearrangement decay modes $\eta_c\phi$ and $D_s^*\bar{D}_s$. The channels $J/\psi\eta^{\prime}$, $J/\psi\eta$, and $D_s^{*+}D_s^{*-}$ are suppressed. This tetraquark has similar properties with the second lowest one. For the highest state which is around 4.3 GeV, it mainly decays into $\eta_c\phi$ and $D_s^{*+}D_s^{*-}$ with a ratio $\Gamma(\eta_c\phi):\Gamma(D_s^{*+}D_s^{*-})\simeq0.2$. So far, no exotic states can be assigned as the $1^{+-}$ $cs\bar{c}\bar{s}$ tetraquarks. Whether such states exist or not needs to be answered by future measurements.

%%%%%%%%%%%%%%%%%%%%%%%%%%%%%%%%%%%%%%%%%%%
\section{Discussions and summary}\label{sec4}
%%%%%%%%%%%%%%%%%%%%%%%%%%%%%%%%%%%%%%%%%%%

The above results used the assignment that the $X_1(4140)$ is the lower $1^{++}$ $cs\bar{c}\bar{s}$ tetraquark. Now we move on to the case using the mass and width of $X(4274)$ rather than $X_1(4140)$ as inputs. Assuming that the $X(4274)$ with mass $4286^{+8}_{-9}$ MeV and width $51\pm 7$ MeV \cite{ParticleDataGroup:2022pth} corresponds to the higher $1^{++}$ $cs\bar{c}\bar{s}$ tetraquark, all the tetraquark masses in Table \ref{mass} would be 27.6 MeV lower. Table \ref{decay2} lists width results we get. All of them are smaller than those in Table \ref{decay1}. In this case, the mass of $X_1(4140)$ is perfectly consistent with the updated value for the LHCb measurement \cite{LHCb:2021uow}, but the width is much smaller. The estimated mass of the lowest $0^{++}$ state is about 45 MeV smaller than the measured value of $X(3960)$ \cite{LHCb:2022vsv}. The obtained width ($19.4$ MeV) is also smaller than the measured value, $43^{+15}_{-15}$ MeV. The second lowest $0^{++}$ state is about 82 MeV below the measured mass of $X_0(4140)$ and its width is smaller than the measured one, which makes the interpretation of $X_0(4140)$ as a $cs\bar{c}\bar{s}$ less reliable. The highest $0^{++}$ state has a mass closer to $X(4350)$ than the previous case, but the width is still larger than the measured value. Comparing the possible tetraquark interpretations in the case using the LHCb results for the $X_1(4140)$ obtained in Ref. \cite{LHCb:2016axx} and the case using the $X(4274)$ as the reference state, one concludes that the former case has a better description than the latter case .

\begin{table}[htbp]\scriptsize
	\caption{Rearrangement decays for the $cs\bar{c}\bar{s}$ states by assigning the $X(4274)$ as the higher $1^{++}$ $cs\bar{c}\bar{s}$ tetraquark. The numbers in the parentheses are ($100|{\cal M}|^{2}/\mathcal{C}^{2},\Gamma$ ) where the coupling parameter $\mathcal{C}$ is extracted from the width of $X(4274)$ (51 MeV \cite{ParticleDataGroup:2022pth}). The partial width $\Gamma$ and total width $\Gamma_{sum}$ are given in units of MeV.}\label{decay2}
	\begin{tabular}{c|c|ccccc|c}\hline\hline
		$J^{PC}$&Mass&\multicolumn{5}{c}{Channels}&$\Gamma_{sum}$\\
		\hline
		&&$J/ \psi \phi$&$D_s^{*+} D_s^{*-}$&&&\\
		$2^{++}$&$\left[\begin{array}{c}4289.3\\4267.0\end{array}\right]$&$\left[\begin{array}{c}(83.4, 35.6)\\(16.6, 6.7)\end{array}\right]$&$\left[\begin{array}{c}(47.5, 14.3)\\(52.5, 12.9)\end{array}\right]$&&&&$\left[\begin{array}{c}49.9\\19.6\end{array}\right]$\\
		&&$J/ \psi \phi$&$( D_s^{*+} D_s^- - D_s^+ D_s^{*-} ) /\sqrt{2}$&&&\\
		$1^{++}$&$\left[\begin{array}{c}4286.0\\4118.8\end{array}\right]$&$\left[\begin{array}{c}(99.8, 42.3)\\(0.2, 0.0)\end{array}\right]$&$\left[\begin{array}{c}(8.2, 8.7)\\(91.8, 45.2)\end{array}\right]$&&&&$\left[\begin{array}{c}51.0\\45.2\end{array}\right]$\\
		&&$J/ \psi \phi$&$\eta_c \eta^{\prime}$&$\eta_c \eta$&$D_s^{*+} D_s^{*-}$&$ D_s^+D_s^-$\\
		$0^{++}$&$\left[\begin{array}{c}4345.0\\4221.4\\4050.9\\3910.6\end{array}\right]$&$\left[\begin{array}{c}(57.1, 27.5)\\(39.5, 13.5)\\(3.1, $-$)\\(0.3, $-$)\end{array}\right]$&$\left[\begin{array}{c}(0.0, 0.0)\\(0.8, 0.4)\\(18.0, 6.6)\\(34.0,  $-$)\end{array}\right]$&$\left[\begin{array}{c}(0.0, 0.0)\\(0.7, 0.6)\\(16.0, 11.5)\\(30.3, 19.4)\end{array}\right]$&$\left[\begin{array}{c}(52.8, 21.2)\\(42.7, $-$)\\(3.8, $-$)\\(0.8, $-$)\end{array}\right]$&$\left[\begin{array}{c}(0.1, 0.1)\\(2.3, 1.5)\\(49.2, 21.3)\\(48.4, $-$)\end{array}\right]$&$\left[\begin{array}{c}48.9\\16.0 \\39.5 \\19.4 \end{array}\right]$\\
		&&$J/ \psi \eta^{\prime}$&$J/ \psi \eta$&$\eta_c \phi$&$D_s^{*+} D_s^{*-}$&$( D_s^{*+} D_s^- + D_s^+ D_s^{*-} )/\sqrt{2}$\\
		$1^{+-}$&$\left[\begin{array}{c}4280.4\\4197.5\\4139.2\\4062.1\end{array}\right]$&$\left[\begin{array}{c}(0.9, 0.4)\\(3.0, 1.2)\\(2.2, 0.7)\\(46.9, 4.4)\end{array}\right]$&$\left[\begin{array}{c}(0.8, 0.6)\\(2.7, 1.9)\\(1.9, 1.3)\\(41.7, 26.3)\end{array}\right]$&$\left[\begin{array}{c}(8.5, 4.7)\\(36.8, 17.4)\\(54.5, 22.0)\\(0.2, 0.1)\end{array}\right]$&$\left[\begin{array}{c}(97.7, 27.4)\\(1.6, $-$)\\(0.1, $-$)\\(0.6, $-$)\end{array}\right]$&$\left[\begin{array}{c}(0.2, 0.2)\\(23.4, 19.5)\\(49.6, 30.0)\\(26.8, $-$)\end{array}\right]$&$\left[\begin{array}{c}33.3\\39.9 \\53.9 \\30.8 \end{array}\right]$\\
		\hline\hline
	\end{tabular}
\end{table}

Up to now, we considered only one case for the mass and width of $X_1(4140)$ which are taken from Ref. \cite{LHCb:2016axx}. We may also adopt the PDG values \cite{ParticleDataGroup:2022pth} or updated LHCb values \cite{LHCb:2021uow} as inputs. Table \ref{decay-PDG} and \ref{decay-new} show the obtained results in these two cases, respectively. The masses using the PDG values are the same as those in the last section, but the widths are much narrower. The width of the higher $1^{++}$ state is at least 26 MeV smaller than the PDG result for the $X(4274)$. Although the width of the highest $0^{++}$ state is compatible with that of $X(4350)$, the width of the (second) lowest $0^{++}$ state is at least 20 (34) MeV smaller than that of $X(3960)$ ($X_0(4140)$). Therefore, the tetraquark picture using the PDG values is also not good as the case considered in the last section. In the case using the updated LHCb values, the masses are approximately equal to those in the case using the $X(4274)$ as the reference state, but the widths are much larger. The feature of width leads to the unacceptable interpretation for the $X(4274)$, $X(3960)$, $X_0(4140)$, and $X(4350)$ as $cs\bar{c}\bar{s}$ tetraquarks.

\begin{table}[htbp]\scriptsize
	\caption{Rearrangement decays for the $cs\bar{c}\bar{s}$ states by assigning the $X_1(4140)$ as the lighter  $1^{++}$ $cs\bar{c}\bar{s}$ tetraquark. The numbers in the parentheses are ($100|{\cal M}|^{2}/\mathcal{C}^{2},\Gamma$ ) where the coupling parameter $\mathcal{C}$ is extracted from the PDG width of $X_1(4140)$ (19 MeV \cite{ParticleDataGroup:2022pth}). The partial width $\Gamma$ and total width $\Gamma_{sum}$ are given in units of MeV.}\label{decay-PDG}
	\begin{tabular}{c|c|ccccc|c}\hline\hline
		$J^{PC}$&Mass&\multicolumn{5}{c}{Channels}&$\Gamma_{sum}$\\
		\hline
		&&$J/ \psi \phi$&$D_s^{*+} D_s^{*-}$&&&\\
		$2^{++}$&$\left[\begin{array}{c}4316.9\\4294.6\end{array}\right]$&$\left[\begin{array}{c}(83.4, 12.3)\\(16.6, 2.3)\end{array}\right]$&$\left[\begin{array}{c}(47.5, 5.5)\\(52.5, 5.3)\end{array}\right]$&&&&$\left[\begin{array}{c}17.8\\7.7\end{array}\right]$\\
		&&$J/ \psi \phi$&$( D_s^{*+} D_s^- - D_s^+ D_s^{*-} ) /\sqrt{2}$&&&\\
		$1^{++}$&$\left[\begin{array}{c}4313.6\\4146.5\end{array}\right]$&$\left[\begin{array}{c}(99.8, 14.6)\\(0.2, 0.0)\end{array}\right]$&$\left[\begin{array}{c}(8.2, 3.0)\\(91.8, 19.0)\end{array}\right]$&&&&$\left[\begin{array}{c}17.6\\19.0\end{array}\right]$\\
		&&$J/ \psi \phi$&$\eta_c \eta^{\prime}$&$\eta_c \eta$&$D_s^{*+} D_s^{*-}$&$ D_s^+ D_s^-$\\
		$0^{++}$&$\left[\begin{array}{c}4372.6\\4249.0\\4078.5\\3938.2\end{array}\right]$&$\left[\begin{array}{c}(57.1, 9.4)\\(39.5, 4.9)\\(3.1, $-$)\\(0.3, $-$)\end{array}\right]$&$\left[\begin{array}{c}(0.0, 0.0)\\(0.8, 0.2)\\(18.0, 2.4)\\(34.0, $-$)\end{array}\right]$&$\left[\begin{array}{c}(0.0, 0.0)\\(0.7, 0.2)\\(16.0, 3.8)\\(30.3, 6.5)\end{array}\right]$&$\left[\begin{array}{c}(52.8, 7.5)\\(42.7, 2.6)\\(3.8, $-$)\\(0.8, $-$)\end{array}\right]$&$\left[\begin{array}{c}(0.1, 0.0)\\(2.3, 0.5)\\(49.2, 7.6)\\(48.4, 0.8)\end{array}\right]$&$\left[\begin{array}{c}16.9\\8.3\\13.8 \\7.3 \end{array}\right]$\\
		&&$J/ \psi \eta^{\prime}$&$J/ \psi \eta$&$\eta_c \phi$&$D_s^{*+} D_s^{*-}$&$( D_s^{*+} D_s^- + D_s^+ D_s^{*-} )/\sqrt{2}$\\
		$1^{+-}$&$\left[\begin{array}{c}4308.0\\4225.1\\4166.9\\4089.8\end{array}\right]$&$\left[\begin{array}{c}(0.9, 0.1)\\(3.0, 0.4)\\(2.2, 0.2)\\(46.9, 3.1)\end{array}\right]$&$\left[\begin{array}{c}(0.8, 0.2)\\(2.7, 0.6)\\(1.9, 0.4)\\(41.7, 8.7)\end{array}\right]$&$\left[\begin{array}{c}(8.5, 1.6)\\(36.8, 5.9)\\(54.5, 7.7)\\(0.2, 0.0)\end{array}\right]$&$\left[\begin{array}{c}(97.7, 10.7)\\(1.6, 0.0)\\(0.1, $-$)\\(0.6, $-$)\end{array}\right]$&$\left[\begin{array}{c}(0.2, 0.1)\\(23.4, 6.9)\\(49.6, 11.6)\\(26.8, 2.1)\end{array}\right]$&$\left[\begin{array}{c}12.7\\13.9\\20.0\\14.0\end{array}\right]$\\
		\hline\hline
	\end{tabular}
\end{table}

\begin{table}[htbp]\scriptsize
	\caption{Rearrangement decays for the $cs\bar{c}\bar{s}$ states by assigning the $X_1(4140)$ as the lighter $1^{++}$ $cs\bar{c}\bar{s}$ tetraquark. The numbers in the parentheses are ($100|{\cal M}|^{2}/\mathcal{C}^{2},\Gamma$) where the coupling parameter $\mathcal{C}$ is extracted from the updated LHCb width of $X_1(4140)$ (162 MeV \cite{LHCb:2021uow}). The partial width $\Gamma$ and total width $\Gamma_{sum}$ are given  in units of MeV.}\label{decay-new}
	\begin{tabular}{c|c|ccccc|c}\hline\hline
		$J^{PC}$&Mass&\multicolumn{5}{c}{Channels}&$\Gamma_{sum}$\\
		\hline
		&&$J/ \psi \phi$&$D_s^{*+}D_s^{*-}$&&&\\
		$2^{++}$&$\left[\begin{array}{c}4288.4\\4266.1\end{array}\right]$&$\left[\begin{array}{c}(83.4, 128.8)\\(16.6, 24.2)\end{array}\right]$&$\left[\begin{array}{c}(47.5, 51.5)\\(52.5, 46.4)\end{array}\right]$&&&&$\left[\begin{array}{c}180.3\\70.6\end{array}\right]$\\
		&&$J/ \psi \phi$&$( D_s^{*+} D_s^- - D_s^{+} D_s^{*-} ) /\sqrt{2}$&&&\\
		$1^{++}$&$\left[\begin{array}{c}4285.1\\4118.0\end{array}\right]$&$\left[\begin{array}{c}(99.8, 152.9)\\(0.2, 0.0)\end{array}\right]$&$\left[\begin{array}{c}(8.2, 31.5)\\(91.8, 162.0)\end{array}\right]$&&&&$\left[\begin{array}{c}184.3\\162.0\end{array}\right]$\\
		&&$J/ \psi \phi$&$\eta_c \eta^{\prime}$&$\eta_c \eta$&$D_s^{*+}D_s^{*-}$&$ D_s^+ D_s^-$\\
		$0^{++}$&$\left[\begin{array}{c}4344.1\\4220.5\\4050.0\\3909.7\end{array}\right]$&$\left[\begin{array}{c}(57.1, 99.6)\\(39.5, 48.6)\\(3.1, $-$)\\(0.3, $-$)\end{array}\right]$&$\left[\begin{array}{c}(0.0, 0.1)\\(0.8, 1.6)\\(18.0, 23.9)\\(34.0, $-$)\end{array}\right]$&$\left[\begin{array}{c}(0.0, 0.1)\\(0.7, 2.1)\\(16.0, 41.8)\\(30.3, 70.3)\end{array}\right]$&$\left[\begin{array}{c}(52.8, 76.5)\\(42.7, $-$)\\(3.8, $-$)\\(0.8, $-$)\end{array}\right]$&$\left[\begin{array}{c}(0.1, 0.4)\\(2.3, 5.4)\\(49.2, 77.0)\\(48.4, $-$)\end{array}\right]$&$\left[\begin{array}{c}176.7\\57.7 \\142.7 \\70.3 \end{array}\right]$\\
		&&$J/ \psi \eta^{\prime}$&$J/ \psi \eta$&$\eta_c \phi$&$D_s^{*+} D_s^{*-}$&$( D_s^{*+} D_s^- + D_s^+ D_s^{*-} )/\sqrt{2}$\\
		$1^{+-}$&$\left[\begin{array}{c}4279.5\\4196.6\\4138.4\\4061.3\end{array}\right]$&$\left[\begin{array}{c}(0.9, 1.5)\\(3.0, 4.3)\\(2.2, 2.4)\\(46.9, 15.1)\end{array}\right]$&$\left[\begin{array}{c}(0.8, 2.0)\\(2.7, 6.8)\\(1.9, 4.7)\\(41.7, 95.3)\end{array}\right]$&$\left[\begin{array}{c}(8.5, 16.9)\\(36.8, 62.7)\\(54.5, 79.4)\\(0.2, 0.2)\end{array}\right]$&$\left[\begin{array}{c}(97.7, 98.7)\\(1.6, $-$)\\(0.1, $-$)\\(0.6, $-$)\end{array}\right]$&$\left[\begin{array}{c}(0.2, 0.9)\\(23.4, 70.3)\\(49.6, 107.8)\\(26.8,  $-$)\end{array}\right]$&$\left[\begin{array}{c}119.9\\144.2 \\194.2 \\110.6 \end{array}\right]$\\
		\hline\hline
	\end{tabular}
\end{table}

Now we take a look at the width ratios mentioned in the last section. When comparing such ratios between the above mentioned four cases, one finds that the width ratio of a tetraquark is affected mainly by whether the tetraquark has the same channels. When the decay channels are the same in these cases, the width ratios are not affected much. When a channel is kinematically forbidden in some case, the ratio is changed accordingly. The involved $cs\bar{c}\bar{s}$ tetraquarks are the highest $0^{++}$, the second lowest $0^{++}$, and the highest $1^{+-}$ states.

From above discussions, the calculated masses and widths of $cs\bar{c}\bar{s}$ tetraquark states by using the reference state $X_1(4140)$ whose mass and width are determined in Ref. \cite{LHCb:2016axx} are more reasonable than other cases. Since the input width of $X_1(4140)$ still has  large uncertainty, the obtained tetraquark widths may be updated. As a model calculation to understand the properties of the observed exotic states, the present study considered only the $cs\bar{c}\bar{s}$ component. In fact, a physical charmonium-like state is probably a mixture of $c\bar{c}$, $cn\bar{c}\bar{n}$ ($n=u,d$), and $cs\bar{c}\bar{s}$ components. The possible assignments discussed in the last section may be improved once the mixture configuration could be considered. In that case, one would probably find appropriate positions for more states like the $X(3930)$ \cite{LHCb:2020bls,LHCb:2020pxc}.

In a previous study \cite{Wu:2018xdi}, we presented the $K$ factors for various $cs\bar{c}\bar{s}$ tetraquark states. From the results, we argued that the highest $2^{++}$, the highest $1^{++}$, and the second highest $0^{++}$ state are probably more stable than other states. Form table \ref{decay1}, one sees that the estimated decay widths do not always satisfy this feature. The reason is that the decay width of a tetraquark is affected by the coupling matrix element, the phase space, and the number of decay channels, while the $K$ factors are just directly related to the coupling matrix elements \cite{Cheng:2020nho}.

To summarize, we have studied properties of the compact $cs\bar{c}\bar{s}$ tetraquark states in the present work. The masses and rearrangement decay widths are estimated with the assumption that the $X(4140)$ is the lower $1^{++}$ $cs\bar{c}\bar{s}$ tetraquark. Our results show that the recently reported state $X(3960)$ announced by the LHCb Collaboration \cite{LHCb:2022vsv} could be assigned as the lowest $0^{++}$ $cs\bar{c}\bar{s}$ tetraquark and the $X(4350)$ observed by Belle \cite{Cheng-Ping:2009sgk} as the highest $0^{++}$ tetraquark. Our results also suggest that the $X_0(4140)$ may be a candidate of the second lowest $0^{++}$ $cs\bar{c}\bar{s}$ tetraquark. The ratios between partial widths of dominant channels for these announced states are predicted. If all the compact $cs\bar{c}\bar{s}$ tetraquarks exist, besides these five candidates, seven states are still awaiting to be observed. Four of them have quantum numbers $J^{PC}=1^{+-}$, two of them have $J^{PC}=2^{++}$, and one of them has $J^{PC}=0^{++}$. Possible finding channels for them are presented. Hopefully, future experimental data can test the predictions.

%%%%%%%%%%%%%%%%%%%%%%%%%%%%%%%%
\section*{Acknowledgments}
%%%%%%%%%%%%%%%%%%%%%%%%%%%%%%%%

This project was supported by the National Natural Science Foundation of China (Nos. 12235008, 12275157, and 11905114) and the Shandong Province Natural Science Foundation (ZR2023MA041).

\end{document}